\documentclass[12pt,a4paper]{article}

\newcommand{\be}{\begin{eqnarray}}
\newcommand{\ee}{\end{eqnarray}}
\newcommand{\bi}{\bibitem}

\newcommand{\rar}{\rightarrow}
\newcommand{\lrar}{\leftrightarrow}

\begin{document}

\begin{center}
CP VIOLATION IN COSMOLOGY \vspace{0.3cm}\\
A.D. Dolgov\\
\it{INFN, Ferrara, Italy\\
and\\
ITEP, Moscow, Russia.}
\end{center}

\begin{abstract}
Cosmological implications of asymmetry between particles and antiparticles are 
reviewed. Three possible mechanisms of CP-violation in cosmology are described.
General features of kinetics of generation of cosmological charge asymmetry 
are discussed in detail. In particular, the cyclic balance condition, which 
plays the same role in time non-invariant theory as detailed balance does in 
T-invariant case, is derived. Several scenarios of baryogenesis are described 
with an emphasis on CP-violation mechanisms. Production of cosmic antimatter 
and a possibility of its ``living'' in our neighborhood is discussed.   
\end{abstract}

\section{Introduction}

In these three lectures I am going to discuss 
the cosmological impact of the
asymmetry between particles and antiparticles and the role that C and
CP violations play in creating the universe in the present suitable for
life form. Seemingly, the breaking of C and CP symmetries are necessary for
our existence but this happens to be true only in the simplest versions 
of the theory. There are realistic and natural scenarios according to 
which charge asymmetric universe can be created even in the case of
C and CP conserving fundamental interactions. We will consider
different possible mechanisms of breaking of C and CP symmetries
in cosmology, some of which are the usual ones created by complex couplings 
or mass matrices, while the others may be operative only in the early 
universe and disappear today. Different models of creation of cosmological
baryon, or any other charge asymmetry will be presented.
A special attention will be paid to physical kinetics 
in the case of a broken time reversal invariance
and to some general properties of elementary processes leading to a
generation of cosmological charge asymmetry.

The generally accepted mechanism of generation of a cosmological
charge asymmetry is based on three famous Sakharov's 
conditions~\cite{ads}:\\
1. Non-conservation of (baryonic) charge.\\
2. Breaking of symmetry between particles and antiparticles, both
C and CP.\\
3. Deviation from thermal equilibrium.\\
We will see in what follows that none of these conditions is
obligatory~\cite{ad-baryo-rev} but in the simplest versions of the
baryogenesis scenarios they are usually assumed to be fulfilled.

As is known from different pieces of astronomical data, the universe,
at least in our neighborhood, is strongly charge asymmetric: it is populated
only with particles, while antiparticles are practically absent. A small
number of the observed antiprotons or positrons in cosmic rays can be
explained by their secondary origin through particle collisions
or maybe by annihilation of dark matter. 
Any macroscopically large antimatter domains or objects (anti-stars,
anti-planets or gaseous clouds of antimatter), if exist, should be
quite rare. As we see in what follows, there are plenty of baryogenesis
scenarios which predict either charge symmetric universe at very large
scales or, even more surprising, an admixture of antimatter
in our vicinity with rather large amount but still compatible with 
observational restrictions. 

The content of these lectures is the following.
In sec. 2 we will see that the baryon asymmetry of the universe
was certainly generated dynamically and was not a result of a
charge asymmetric initial state. In sec. 3 the Sakharov conditions 
are considered. Kinetic equations with broken detailed balance
and, possibly, without CPT are analyzed in sec. 4. General 
features of the observed baryon asymmetry and possible existence
of cosmological antimatter are discussed in sec. 5. Three mechanisms 
of C and CP violation which may be operative in cosmology are
presented in sec. 6. In secs. 7 and 8 two most conservative
scenarios of baryogenesis: electroweak and heavy particle decays
are respectively described. In sec. 9 several more speculative models of
baryogenesis are presented. In particular, creation of cosmic
antimatter and testable consequences of these models are discussed.

\section{Dynamical or accidental?}

The first question to address is 
whether the observed predominance of matter over
antimatter is dynamical or accidental? The former should be
generated by some physical processes in the early universe starting
from a rather arbitrary initial state, while the latter could be
created by proper
initial conditions? Such a question could be sensible a quarter
of century ago, but now with established inflationary cosmology, the
answer may be only that any cosmological charge asymmetry should have 
been generated dynamically~\cite{ad-baryo-rev,ad-ms-yz}.

First of all I would like to stress that inflation is practically an
``experimental'' fact. We do not know any other way to 
create the universe with the observed properties (for review 
on inflation see e.g.~\cite{inf-rev}).
Inflationary cosmology naturally explains:\\
1. Flatness of the present-day universe, 
$\Omega =\rho/\rho_c =1$, here $\rho$ is the average cosmological 
mass/energy density and $\rho_c = 3H^2 m_{Pl}^2/8\pi$ is the critical
energy density. Without inflation 
the adjustment of $\Omega$ should be at the level ${ 10^{-60}}$ at the 
Planck era or about $ 10^{-15}$ during primordial nucleosynthesis.\\
2. The well known problems of homogeneity, isotropy, and horizon, which
created head-ache in frameworks of the
old Friedman cosmology, are uniquely and 
beautifully solved.\\
3. Inflation presents a natural mechanism of generation of small density
perturbations with practically flat spectrum in agreement with 
observations. The only known competing mechanism of creation of density
perturbations by topological defects
seems to be ruled out by the angular spectrum of the Cosmic Microwave
background radiation (CMBR)~\cite{cmb-top-dfcts}. (Maybe better to
say that topological defects cannot be the dominant source of density
perturbations.)

To fulfill all these jobs inflation should last at least 70 Hubble times.
During this period the energy density must remain (approximately)
constant, otherwise expansion would not be an exponential one but only
a power law. Indeed the Hubble parameter is:
\be
\dot a /a \equiv H \sim \sqrt {\rho}/m_{Pl}
\label{hubble}
\ee 
The scale factor would rise exponentially if $\rho = const$. On the other
hand, if there is a conserved charge the energy density
cannot stay constant. The density of any conserved, in particular baryonic, 
charge scales as 
\be
B\sim 1/a^3
\label{b-of-a}
\ee
and the energy density associated with this charge evolves as
\be
\rho_B \sim 1/a^n,
\label{rho-B}
\ee
where $n=4$ for relativistic matter and $n=3$ for nonrelativistic matter.

According to observations, the ratio of the baryonic charge density
to the number density of photons is 
\be
\beta = B/N_\gamma \approx 6\cdot 10^{-10},
\label{beta}
\ee
Let us assume now that we go backward in time and reach inflationary
era. As we mentioned above, at inflationary period the energy density of 
matter must be (approximately)
constant. For conserved baryons
it could be true if baryons were sub-dominant in the energy 
density. However, it is clear that according to  eqs. (\ref{rho-B})
and (\ref{beta})
the energy density associated with conserved baryonic charge 
should rise with decreasing time as $ \rho_B \sim \exp (-4Ht)$ and 
for $Ht = 4-5$ the sub-dominant baryons would become dominant. Thus the
energy density could not be maintained constant and inflation would
be destroyed after too short duration, $Ht \approx 5 \ll 70$.

Hence successful inflation is incompatible with baryonic charge  
conservation and baryon asymmetry must be generated dynamically.

\section{Three Sakharov's conditions - discussion}

\subsection{Non-conservation of baryons}

In 1967, when Sakharov~\cite{ads} first presented 
his work on baryogenesis, a 
non-conservation of baryonic charge was the weakest point. No theory
demanded that and, clearly, there were no experimental data in support
of this hypothesis. Today direct experimental proof that baryons are
not conserved are still missing but there is great theoretical progress:
it is shown that grand unification and even the standard electroweak 
theory demand nonconservation of baryonic charge. Another very interesting
possibility is that gravity itself, most probably, breaks all global
charges and, in particular, baryonic charge~\cite{hawk-bs,zel-bs,zel-decay}. 

The only ``experimental piece of data'' in favor of non-conservation 
of $B$ is our universe:
{\it we exist, ergo baryons are not conserved.}
A half of century ago the same piece of knowledge led to the
opposite conclusion: {\it we exist, ergo baryons are conserved.}
This surely shows a necessity of a theory for interpretation of even
very clear experimental data.

\subsection{C and CP violation}

In contrast to non-conservation of baryons, the breaking of CP-symmetry,
was discovered in 1964 in direct experiment~\cite{cronin}.
It was a great surprise
to community, despite of parity violation broken in 1956~\cite{parity}
and almost simultaneous realization that charge symmetry, C, is also
broken. In an attempt to preserve symmetry between matter and antimatter
it was suggested~\cite{cp} that combined parity, CP, i.e. the 
transformation from particles to mirror reflected antiparticles is a 
rigorous symmetry of Nature. It happened not to be the case. The particles
and antiparticles are really different and after this discovery ``life in
the universe became possible''. (As we will see in what follows, 
this is not exactly so because the generation
of cosmological charge asymmetry and thus creation of suitable for life
universe is possible even without CP-violation in particle physics, but
still some asymmetric conditions must be created dynamically even
in a charge symmetric theory.)

A natural question may arise at this stage: if breaking of C-symmetry
only may be sufficient for generation of cosmological charge asymmetry?
The answer is negative by the following ``global'' (presented here)
and ``local'' (presented later, eq. (\ref{sum-spins})) 
arguments. Let us assume that the universe
is initially charge symmetric i.e. she is in C eigenstate:
\be
C|u\rangle = \eta |u\rangle
\label{c-universe}
\ee
where $|\eta|=1$. This means, in particular, that the universe 
has all zero charges. May some
non-zero charge, e.g. ${ B}$, be generated dynamically if CP is conserved
but C is violated?

Let us assume first that even C is conserved. Then it is evident that
no charge asymmetry could be created. The formal arguments go as follows.
Since C is conserved the operator of C-transformation commutes with the
Hamiltonian:
\be
\left[ C, {\cal H}\right] = 0
\label{c-h}
\ee
Time evolution of baryonic charge density ${ B}$ is governed by the
equation:
\be
B(t) = \langle u | e^{-i {\cal H}t} J^B_0 e^{i{\cal H}t} |u\rangle.
\label{B-of-t}
\ee
Let us insert now the unity operator ${ I = C^{-1} C}$ into 4 places above:
\be
B(t) =\langle u| I
 e^{-i{\cal H}t} I J^B_0 I e^{i{\cal H} t} I
| u \rangle = - B(t) = 0
\label{c-b-of-t}
\ee
taken that baryonic current is a C-odd operator, 
$ C J^B_0 C^{-1} = -J_0^B $.
Thus in C-conserving theory B or any other charge cannot be generated.

The same arguments with ${ CP}$ instead of ${ C}$ leads to the
conclusion that with conserved ${ CP}$ no charge asymmetry could be 
generated if {the universe is an eigenstate of CP:}
\be
CP |u\rangle = \eta |u\rangle
\label{cp-univ}
\ee
However, e.g. in a globally rotating universe charge asymmetry might be
generated even if CP is conserved: {\it global rotation could be
transformed into baryonic charge!?} 
{To this end new interactions, almost surely long range and
possibly pathological, are needed.}

If cosmological charge asymmetry could be generated only in local 
processes, as in out-of-equilibrium decays of heavy
particles, then one can easily see that in CP-invariant situation
the partial decay rates are equal for charge conjugated channels
and resulting asymmetry is zero - it is discussed in detail below.
However, if decaying particles are globally polarized with respect 
to some fixed axis, then a charge asymmetry may be produced even
in CP invariant theory. 

Out of the three symmetries, P, C, and T, each of which was believed to 
be true in the first part of the XX century, only their product, 
CPT, survived by now. And there is a good reason for that: while 
breaking of any of these three individual symmetries does not contradict
``sacred'' physical principles, CPT invariance is
the only symmetry with solid theoretical
justification, the famous CPT-theorem~\cite{cpt}. It can be
proved that Lorentz-invariant theory with
canonical relation between spin and statistics is automatically 
invariant with respect to CPT transformation. 
The situation with C, P, and T transformations demonstrates
the validity of the principle ``all what is not forbidden is allowed 
and exists''. At the present time there are no serious
data indicating to
breaking of CPT-invariance.

A simple consequence of CPT invariance, namely that
breaking of CP leads
simultaneously to breaking of T is important for kinetics of
charge asymmetry generation, as we see in sec. \ref{s-equil}.

Despite of CPT-theorem, some models without CPT are considered, e.g.
for explanation of neutrino anomalies or just 
``why not'?''~\cite{break-cpt}. Keeping this in mind we will
discuss the generation of cosmological
charge asymmetry with broken CPT as well.

\subsection{ Thermal equilibrium}

In thermal equilibrium particle number densities in phase space
are given by the expressions:
\be
f_{eq} = \left[ \exp \left(\frac{E-\mu}{T}\right) \pm 1\right]^{-1}
\label{f-eq}
\ee
where signs $\pm$ refer to Fermi and Bose statistics and 
$E = \left( p^2 + m^2\right)^{1/2}$ is particle energy.
In equilibrium chemical potential for a 
non-conserved charge must vanish: ${{\mu =0}}$ (see below).
Due to CPT-invariance the masses of particles and antiparticles
are equal, ${ m=\bar m}$ and thus:
\be
n = \bar n = \int \frac{d^3 p}{(2\pi)^3}\,\, f_{eq}
\label{n=nbar}
\ee 
By definition equilibrium distributions are the solutions of
the kinetic equation:
\be
\frac{df}{dt} = I_{coll} [f],
\label{kin-eq}
\ee
which annihilate the collision integral:
\be
I_{coll} [f_{eq}]=0.
\label{i-coll=0}
\ee
We will check later that
${ f_{eq}}$ presented above, eq.~(\ref{f-eq}), indeed annihilate collision 
integrals due to {conservation of energy}:
\be
\sum E_{in} = \sum E_{fin}
\label{e-conserv}
\ee
and of chemical potentials:
\be
\sum \mu_{in} = \sum \mu_{fin},
\label{mu-conserv}
\ee
While the first condition should be always true, the second one is
enforced by reactions if they are fast enough to establish equilibrium.
Here sub-indices ``in'' and ``fin'' refer respectively to initial and
final particles.

Now we can check that chemical potentials of baryons vanish
if the processes with non-conservation of B-charge are in equilibrium.
It may happen that baryonic charge is conserved or its non-conservation
is so weak that equilibrium with respect to reactions with 
$\Delta B \neq 0$ is never established. In this case baryonic chemical
potential would not evolve with time and would remain equal to its
initial value. Correspondingly baryonic charge density would be
constant in comoving volume (i.e. in volume which expands together with
the universe, $V\sim a^3$, where $a(t)$ is the
cosmological scale factor).
 
We assume that complete equilibrium with respect to all reactions
is established and check what happens with chemical potentials.
First, as is well known, chemical potential of photons in equilibrium
is zero. Indeed, the number of photons is not conserved and the following
reactions with different number of photons in the final state are
possible:
\be
q +\bar q \rar 2\gamma,\,\,\, q+\bar q \rar 3\gamma 
\label{q-into-gamma}
\ee
In equilibrium:
\be
 \mu_q + \mu_{\bar q} = 2\mu_\gamma = 3\mu_{\gamma} =0
\label{mu-gamma}
\ee
We see, in particular, that in equilibrium chemical potential of particles 
and antiparticles are opposite:
\be
\mu_q = -\mu_{\bar { q}}
\label{mu-q}
\ee
If there is an excess of $q$ over $\bar q$ or vice versa, it
is described by a non-vanishing $\mu_q$. 

If B-nonconserving processes are in equilibrium, then e.g. the reaction
(or any other with different baryonic number in the final and
initial states)
\be
q+q+q \lrar \bar q + \bar q +\bar q
\label{3q}
\ee
leads to ${ 3\mu_q = 3 \mu_{\bar q}}$
and simultaneously the condition (\ref{mu-q}) is valid. Thus:
\be
\mu_q = \mu_{\bar q} =0.
\label{mu-q=0}
\ee

Normally because of large magnitude of the Planck mass,
$m_{Pl} \approx 1.2\cdot 10^{19}$ GeV, 
the cosmological expansion rate
\be
H = \frac{\dot a}{a} \sim  \frac{\sqrt \rho}{m_{Pl}}
\sim  \frac{T^2}{m_{Pl}}
\label{h}
\ee
is much smaller than the reaction rates in the hot primeval plasma and 
because of that deviations from equilibrium are very weak. (In the
equation above $\rho \sim T^4$ is the energy density of cosmological 
plasma with temperature $T$.) Moreover, for massless particles no
deviation from equilibrium is induced by the cosmological expansion. Still
for massive particles some deviation from equilibrium always exist,
though suppressed by the small ratio:
\be
\frac{H}{\Gamma} \sim \frac{T^2}{m_{Pl}\Gamma}
\label{h/gamma}
\ee
where $\Gamma$ is the rate of processes creating equilibrium. 
For heavy particle decays ${ \Gamma \sim \alpha m}$, while for two-body
reactions with massless or light particles
${ \Gamma \sim \alpha^2 T}$, where $\alpha$ is usually
close to 1/100.

In Friedman-Robertson-Walker cosmology kinetic equation has the form:
\be
\frac{df}{dt}= \left( \partial_t + \dot p \partial_p \right) f
= \left(\partial_t - Hp \partial_p \right) f 
\label{kin-eq-cosm}
\ee
since ${ \dot p = - Hp}$.
As we mentioned above, in equilibrium $I_{coll}$ must vanish and so 
must ${ df/dt}$. Let us check if this may be fulfilled 
for massive particles:
\be
 \left(\partial_t - Hp \partial_p \right) 
f_{eq}\left(\frac{E-\mu(t)}{T(t)}\right) = 
\left[-\frac{\dot T}{T}\,\frac{E-\mu}{T} -\frac{\dot \mu}{T}  
-\frac{Hp}{T}\right]  f'_{eq}
\label{dt-massive}
\ee
where prime means derivative with respect to the argument of $f_{eq}$.
The factor in square brackets vanishes if, firstly, 
\be 
\dot \mu = \dot T/T= -H,
\label{mu-dot}
\ee
this can be true, and, secondly, 
\be
E(\dot T /T) = -Hp,
\label{e-dot-t}
\ee
This can be true only for massless particles for which 
$ E=p $. For example cosmic microwave background radiation
(CMBR) has perfect equilibrium spectrum, 
because photons are massless. 
Photons were in equilibrium at high temperatures and even when the 
interactions were switched-off after hydrogen recombination, the 
equilibrium distribution of photons was not distorted by the 
cosmological expansion.

The magnitude of
deviation from equilibrium for massive particles can be estimated
from the kinetic equation with a simplified collision integral
\be
Ha\frac{\partial f}{\partial a} = \Gamma \left(f_{eq} -f \right)
\label{kin-simpl}
\ee
where ${ a\sim 1/T}$ is the
cosmological scale factor. This equation can be
obtained from eq.~(\ref{kin-eq-cosm}) after change of variables
to $a(t)$ and $y=p/a(t)$ instead of $t$ and $p$ and the substitution
into the r.h.s. of eq. (\ref{kin-simpl}) the simplified algebraic 
expression, instead of the exact collision integral.
This approximation is reasonably good for small deviations from
equilibrium, $ f=f_{eq} +\delta f$. In this case it is easy to 
find:
\be
\frac{\delta f}{f_{eq}} \approx \frac{Hm^2}{\Gamma E T}
\approx \frac{m}{\alpha m_{Pl}}  
\label{delta-f}
\ee
The last estimate is obtained for ${\Gamma \sim \alpha m}$. 
The temperature was taken of
the order of the mass of the decaying particle, $T \sim m$. For 
larger $T$ the deviations from equilibrium are small, while for 
smaller $T\ll m$ the equilibrium distribution is Boltzmann suppressed,
$f_{eq} \sim \exp (-m/T)$, and so is $\delta f$, while the relative
deviation from equilibrium can  be large.

Thus for successful baryogenesis either heavy particles are needed or low
decay rate, i.e. small coupling constant, $ \alpha\ll 10^{-2}$, 
i.e. $\alpha$ must be much smaller than the natural value of the 
gauge coupling constant. Thermal equilibrium can be strongly
broken even with light particles ($m\ll m_{Pl}$) in the case of
first order phase transitions when two phases coexisted in the
primeval plasma.

A new possibility to break thermal equilibrium even without 
immediate action of massive particles was suggested in ref.~\cite{bb}.
The authors considered two weakly interacting thermal bathes, each
being an equilibrium one, but with different temperatures. Such a 
case could be realized with mirror matter if temperature of mirror
world was different from ours. The situation is similar to 
the breaking
of equilibrium between electrons/photons and neutrino in the 
standard cosmology after $e^+e^-$-annihilation which led to an
increase of $e\gamma$ temperature but leaves neutrino temperature
practically unchanged. Residual interactions between $e^\pm$ and
neutrinos distorts spectrum of massless neutrinos~\cite{df}.

\section{Baryon asymmetry with broken CPT and validity of standard 
equilibrium distributions in CPT or T violating theories
\label{s-equil}}

Evidently, if CPT is broken and masses of particles and antiparticles
are unequal, then charge asymmetry can be generated in thermal equilibrium,
see e.g.~\cite{ad-yz}.
The difference between the number densities of particles and 
antiparticles is equal to:
\be
\frac{N_B-N_{\bar B}}{N_B} = \int \frac{d^3 p}{(2\pi)^3} \left[f_B (p)
- f_{\bar B}(p) \right]
\label{n-bar-n}
\ee
If equilibrium distributions maintain the same standard
form (\ref{f-eq}), despite CPT breaking, then neglecting possible
chemical potentials, we obtain:
\be
\frac{N_B-N_{\bar B}}{N_B} \approx \frac{\delta m}{T},
\,\,\, {\rm for}\,\,\, m>T,\,\,\, {\rm but}\,\,\,
\delta m <T 
\label{m>T}
\ee
and 
\be
\frac{N_B-N_{\bar B}}{N_B} \approx \frac{\delta m}{T}\,\frac{m}{T} 
\,\,\, {\rm for},\,\, m<T.
\label{m<T}
\ee
Two comments are in order here. \\
1. Care should be taken of electric charge neutrality.
Cosmological electric charge asymmetry must be zero or extremely small.
For a closed universe even a  single excessive electron is not 
allowed. The condition of vanishing electric charge 
demands non-zero chemical potentials of protons (quarks) and electrons
and the results presented above should be modified in a model dependent 
way. \\
2. The results above may be correct only
if the equilibrium distributions are not damaged by
breaking of CPT and/or T invariance.

Let us first discuss whether the 
usual equilibrium distributions remain the same if T-invariance is broken
The problem is related to the fact that the annihilation of collision
integrals by the equilibrium functions given by eq. (\ref{f-eq})
is usually
verified under assumption of detailed balance condition which is
true because of T-invariance. Now we see what happens if T-invariance
is broken. The collision integral has the form:
\be
I_{coll} =\frac{1}{2E_1}\sum_{fin}
\int d\tau'_{in} d\tau_{fin}
 \left[ {|A_{if}|^2} \Pi f_{in} 
\Pi (1\pm f_{fin}) 
 -{|A_{fi}|^2} \Pi f_{fin} \Pi (1\pm f_{in})
\right],
\label{i-coll}
\ee
where particle number 1 is the one whose  evolution is studied, integration
is taken over the phase space of all particles in the final state, 
$d\tau_{fin}$,
and all but 1 particles in the initial state, $d\tau'_{in}$, 
$A_{if}$ and $A_{fi}$ are the amplitudes of initial to final and
final to initial reactions respectively; summation is taken over all
possible final states.

In T-invariant theory the {\it detailed balance} condition is fulfilled: 
\be
{|A_{if}|^2 = |A_{fi}|^2}
\label{det-bal}
\ee
(after some evident change of variables). Hence the amplitudes can
be factored out of the square brackets in eq. (\ref{i-coll}) and
there remains:
\be
\left[\Pi f_{in} \Pi (1\pm f_{fin}) - \Pi f_{in} \Pi (1\pm f_{fin})
\right]
\label{sq-bfrack}
\ee
As we have seen above this expression vanishes for ${ f=f_{eq}}$
because of energy conservation (\ref{e-conserv}) and 
conservation of chemical potentials (\ref{mu-conserv}).

If however T-invariance is broken then the detailed balance
equality (\ref{det-bal}) is not fulfilled and one
should expect $|A_{if}|^2 \neq |A_{fi}|^2$.
A natural suspicion arises in this case if the equilibrium 
distributions may become different in T-noninvariant world?
For the usual equilibrium functions we can rewrite the expression
in square brackets of eq. (\ref{i-coll}) as
\be
I_{coll} \sim \sum_{fin}\Pi f_{in}(1\pm f_{fin}) 
\left( |A_{if}|^2 - |A_{fi}|^2 \right)  
\label{i-coll-no-t}
\ee
The last factor is evidently non-vanishing and one is tempted to 
conclude that if detailed balance condition is invalid, the 
equilibrium distribution are to be modified.

One should remember, however, that T-violation is observable only if 
several 
reaction channels are open. If only one reaction channel is
allowed the amplitude of direct and inverse processes may differ only by
a phase factor and their absolute values are equal. Now we will show that,
though individual terms in sum (\ref{i-coll-no-t}) are non-zero
(detailed balance is violated), their sum vanishes and equilibrium
distributions remain the same as in the standard T-invariant 
theory~\cite{cycle,chang}. 

To see that we will use the unitarity of the
scattering matrix, $S S^\dagger =1$. Separating in the usual
way the unity matrix $I$, $S=I+ iT$,
we find that $T$-matrix, which describes scattering amplitudes,
satisfies:
\be
i\left( T_{if}-T_{fi}^\dagger \right) = 
- \sum_n\, T_{in} T^\dagger_{nf}
= -\sum_n\, T^\dagger_{in} T_{nf}
\label{im-T}
\ee
Here summation over $n$ includes integration over phase space.
$T$-matrix differs from the usual scattering amplitude by some simple
factor.

If T-invariance is broken, still the unitarity of  S-matrix leads, 
instead of detailed balance, to a new condition: 
\be
\sum_k\int d\tau_k \left(|A_{ki}|^2 - |A_{ik}|^2\right) =0
\label{cyclic}
\ee
Here $d\tau_k$ includes Bose/Fermi 
enhancement/suppression factors.
Equation (\ref{cyclic}) ensures vanishing of $I_{coll}$ for 
$ f=f_{eq}$ and thus, even in the absence of detailed balance,
the equilibrium distributions remain the same. 

Equation (\ref{cyclic}) can be called {\it cyclic balance} 
condition~\cite{cycle} because equilibration is now realized
not simply by equality of probabilities of direct and inverse 
reactions but by more complicated cycle of all relevant reactions.

In conclusion of this subsection the following comments may be
useful:\\
1. As is evident from eq. (\ref{im-T}),
in the lowest order $ A_{if}^* = A_{fi}$ and the effects of T-breaking
are unobservable. That's why several intermediate states are to be 
taken into account to produce an additional imaginary part of the 
amplitude created by particle rescattering.\\
2. Full unitarity is not necessary. Normalization of probability
\be
\sum_f w_{if} =1
\label{norm}
\ee
plus CPT invariance are sufficient.\\
3. In the case that CPT is broken, another relation in addition to
(\ref{norm}) is needed:
\be
\sum_f w_{fi} =1
\label{norm-invers}
\ee
to save the standard equilibrium distributions.
4. If conditions 1-3, specified above, are not true, then
small violations of CPT and unitarity could
be strongly enhanced in a system relaxing during long time interval
and the effects of CPT breaking
in kinetics and statistics 
would be {\it large} if simultaneously with CPT 
equations (\ref{norm}) and/or (\ref{norm-invers}) are broken. Maybe 
stationary equilibrium distributions do not exist in this case.

\section{General features of cosmological baryon asymmetry}

Baryons by number are quite rare in the universe. According to
the data there is 1 baryon per 2 billions photons in CMBR,
see eq. (\ref{beta}). While there are about 400 photons per cm$^3$,
there is only 1 proton per 4m$^3$. The number of baryons which
are directly observed is even smaller than that, almost by an
order of magnitude. Majority of cosmic baryons are 
invisible~\cite{invis-b}. Nevertheless, 
their total number density (\ref{beta}) is quite well known 
from two independent
pieces of observations: from abundances of light elements
created during big bang nucleosynthesis (BBN)~\cite{beta-bbn}
and from angular fluctuations of CMBR~\cite{beta-cmbr}.
Both ways of determination of $\beta$ give similar results
and one may speak of another success of the standard cosmological
model but there is an inconsistency between abundances 
of $He^4$ and deuterium which indicate to different values of $\beta$.
Deuterium data rather well agree with $\beta$ found from
CMBR, while helium disagrees with the latter by factor two. 
It is unclear if the data are accurate enough to conclude that
there exists a real disagreement, note only that measurements of
light elements are done at the red-shifts
$z\ll 1$ for helium and $z\sim 1$ for deuterium. On the other hand,
the data on the angular fluctuations of CMBR 
inform us about the universe at $z\approx 10^3$ with
$\beta$ effectively averaged over all the sky. 
Maybe the discrepancy between $^4 He$ and $^2 H$ can be
explained by small fluctuations of the baryon number density
at relatively small spatial scales.
Moreover, according to the observations, the deuterium abundance 
quite strongly fluctuates from point to point on the sky.

The measured small value of $\beta$ is still huge in comparison
with the baryon-to-photon ratio which would be in baryo-symmetric
cosmology. In the universe with locally equal numbers of baryons
and antibaryons we should expect:
\be
 n_B/n_\gamma=n_{\bar B} /n_\gamma \approx 10^{-19}
\label{beta-sym}
\ee
Life in such universe would surely be impossible for us.

Fortunately the universe is not baryo-symmetric at least in our 
neighborhood. We know for sure that the Galaxy is matter dominated
and only a small fraction of antimatter is allowed by observations.
We cannot say if distant galaxies are made of matter or antimatter
but in all the cases when colliding galaxies or galaxies
in the same cloud of intergalactic gas are observed, an absence of 
noticeable annihilation indicates that such galaxies, as well as
the intergalactic gas are dominated by matter (or antimatter).
An absence of noticeable gamma ray line from $\pi^0$ decay
(the latter would come from $\bar p + p \rightarrow \pi$'s) 
allows to conclude that the nearest rich antimatter region
should be away from us at least at 10 Mpc~\cite{steigman}.

A much stronger bound, valid for average baryo-symmetric universe
was derived in ref.~\cite{cdg}. According to the results of these works
the nearest antimatter domain should be at the distance of the order of 
gigaparsec, i.e. comparable to the present day cosmological horizon. 
This bound was obtained from the data on the cosmic gamma ray background.
Proton-antiproton annihilation should be very efficient on the 
boundaries between matter and antimatter domains producing energetic
photons. Naively one would expect that the annihilation
should create an excessive pressure which push matter and antimatter
apart. According to ref.~\cite{cdg}, the picture is opposite.
An excessive pressure is indeed created but at a large distance from
the boundaries where the annihilation products slow down. This produces
an opposite effect of pushing matter and antimatter towards each other
and strongly enhances the annihilation. A study of possible distortions 
of CMBR spectrum to constraint the amount of antimatter leads to a
noticeably weaker bound~\cite{anti-cmbr}. 

It is still an open question if the baryon asymmetry is the same
everywhere in the universe, $\beta = const$, or it may be a function
of space points, $\beta = \beta (x)$. In particular, could the universe
be charge asymmetric locally but charge symmetric globally, i.e. may her
average baryonic charge be zero? As we mentioned above for a globally 
symmetric universe the characteristic scale, $l_B$, of variation of
$\beta$ is close or larger than the present day horizon size. Much more
interesting from the point of view of a discovery of cosmological
antimatter is the possibility of baryo-asymmetric universe dominated
by baryons (or maybe even by antibaryons). In this case astronomically
large clouds of antimatter or compact objects (anti-stars) can be quite
close to us, even in our Galaxy. Compact antimatter objects may
escape observations even if their mass fraction is quite high. Disperse
clouds of antimatter are much better visible.

An answer to the above questions 
and scenarios of creation of relatively rare but astronomically 
large antimatter domains
as well as possibly more abundant compact antimatter
objects, which can be not too far from us,
depends upon the mechanisms of CP violation
realized in cosmology. They are discussed in the following section.

\section{Cosmological CP violation}

\subsection{Explicit CP violation \label{ss-explct}}

In majority of baryogenesis scenarios an explicit CP violation in
the underlying particle theory is assumed. The latter is realized by 
an introduction of complex coupling constants into Lagrangian. Hermitian
conjugation corresponds to transformation from particles to antiparticles  
and for complex couplings the
charge conjugation symmetry becomes broken.
This is the standard way how CP (or C) is broken in particle physics.

This mechanism of CP violation is realized in most popular (though
at different periods) scenarios of baryogenesis (they are reviewed 
e.g. in refs.~\cite{ad-baryo-rev,bs-rev}):\\
{1. GUT baryogenesis.}\\
{2. Electroweak baryogenesis.}\\
{3. Baryo-thru-lepto-genesis.}\\
In all these models, based on explicit CP-violation, $\beta$ is 
an universal ``cosmological constant'' which can be expressed through 
masses and couplings of fundamental particles. (Not confuse this 
``baryonic cosmological constant'' with the usual cosmological constant
which is equivalent to vacuum energy.)

There is no space for cosmic antimatter in this rather dull picture 
but if these scenarios are combined with other models
of cosmological CP-violation a more complicated and interesting
pattern may emerge.

CP-violation is introduced into the minimal standard model (MSM) 
of particle physics, and not only into it, by a complex mass matrix of 
quarks and leptons. The former is called CKM (Cabibbo-Kobayashi-Mascawa)
matrix. The masses become complex because of complex Yukawa
coupling constants of Higgs field to fermions,
$g_{ij} H \bar \psi_i \psi_j$, where $i,j$ are quark flavors and the
interaction of Higgs boson with quarks is not necessarily diagonal in 
the flavor basis. When the Higgs condensate is formed
fermions acquire masses $m_{ij} = g_{ij} \langle H \rangle$.

It is evident that CP-violation in MSM
is absent for two quark families because any phase in $2\times 2$ matrix
$m_{ij}$ can be rotated away by the phase transformation of quark
wave functions, 
\be
q_i\rar \exp (-i\phi_i)\,q_i. 
\label{q-phase}
\ee
The diagonal entries $m_{ii}$ must be real because of hermicity and 
the only off-diagonal one, $m_{12}$, can be always made real by 
transformation (\ref{q-phase}).

At least three quark families are necessary to give rise to observable
CP-violation due to complexity of the quark mass matrix. Indeed 
there can be three phases in the off-diagonal components of $m_{ij}$:
$\phi_{ij}=-\phi_{ji}$ with $i,j = 1,2,3$.
Each of them can be changed by the quark phase rotation as: 
\be
m_{ij} \rar e^{i(\phi_i - \phi_j)} m_{ij} \equiv e^{i\phi_{ij}} m_{ij}
\label{m-ij}
\ee
Here $ \phi_{12} + \phi_{23} + \phi_{31} =0$ and
the phase freedom of the quark wave functions allows
nullify only two phases out of three independent ones.

This feature demonstrates the necessity of three 
quark families for CP-breaking 
and could be an anthropic explanation why these three families are
needed. However, as we see below, CP-violation in the standard model
is extremely weak at high temperatures and baryogenesis is not
efficient enough.
If however a workable model of baryogenesis is found based on 
CP-breaking in the quark sector (with 3 families) we would have an
explanation why there are three and not just one quark family in
Nature.

If {masses} of different up or down quarks are equal, CP violating 
phase can also be {rotated away} because unit matrix is invariant 
with respect to unitary transformation between different
quarks. If the mass matrix is partly diagonal, i.e. some of off-diagonal
entries $m_{ij}$ vanish, 
CP-violation can be {rotated away} as well.

Thus, CP-breaking is proportional to the product of
{the mixing angles} and to
{the mass differences} of all down and all up quarks:
\be
A_- \sim J\,
(m_t^2-m_u^2)(m_t^2-m_c^2)(m_c^2-m_u^2)
(m_b^2-m_s^2)(m_b^2-m_d^2)(m_s^2-m_d^2)/ M^{12}
\label{a-odd}
\ee
where the  Jarlskog determinant is $J\approx
\sin \theta_{12} \sin \theta_{23} \sin \theta_{31} \sin \delta\approx
3\times 10^{-5} \sin\delta$ and $\delta$ is 
a CP-odd phase, while $\theta_{ij}$ are the angles describing the mixing
of flavor states in terms of mass eigenstates. The normalizing mass 
$M$ depends upon the process under consideration. In the case of
electroweak baryogenesis, see sec.~\ref{s-ew}, it is
taken of the order of the electroweak scale, $M\sim 100 $ GeV.

This consideration shows that CP violation could manifest itself
only in high orders of perturbation theory with all quarks 
participating as virtual particles. A diagram of this kind has been
studied in ref.~\cite{12order} where it was argued that CP-violation
appears only in the 12th order of perturbation theory. However, in
a later paper~\cite{ibk-cp} a lower, but still high order, diagram
has been found with logarithmic dependence on the
quark masses which leads
to similar vanishing of the amplitude when masses are equal
but the amplitude is several orders of magnitude larger.

The result above is valid for the Dirac mass matrix, the only type of
the mass matrix which is allowed for charged fermions. For neutral
fermions, e.g. neutrinos, Majorana mass term is also possible:
\be
{\cal L}_M = M_{ij} \nu_i C \nu_j + h.c.,
\label{m-major}
\ee
where $C$ is the operator of charge conjugation. Now all elements of
${ M_{ij}}$ including the diagonal ones may be {complex.}
One can kill three phases in ${ M_{ii}}$ by 3 phase rotations of
$\nu_i$. No freedom is left after that and three phases of
${ M_{12}, M_{23}, M_{31}}$ remain arbitrary. Thus, in the Majorana case
there can be three independent CP-odd phases.

If small neutrino masses are generated by the well known 
seesaw mechanism through mixing of light left-handed neutrinos,
$\nu_L$, with heavy right-handed ones, $L_R$,
the mass matrix can be written as:
\be
{\cal L}_M = M L_R C L_R +m \nu_L C L_R = h.c. 
\label{see-saw}
\ee
After diagonalization of this matrix it will be reduced
to two $3\times 3$ matrices in the light and heavy sectors, each
having 3 independent CP-odd phases.
Evidently the number of CP-odd phases is now six:
3 phases in light $\nu$ sector and 3 phases in heavy $\nu$ sector. 
The phases which may be measured in neutrino oscillations are
not directly related to the phases in heavy $L_R$ decays and low
energy measurements cannot  teach us about CP-violation at
leptogenesis in model independent way.

\subsection{Spontaneous CP violation}

If the Lagrangian is invariant with respect to CP transformation
but the ground states are not, we speak about spontaneous
symmetry breaking. Such a mechanism for CP violation was suggested
in ref.~\cite{spont-cp} in an attempt to save global symmetry between
particles and antiparticles. 

A concrete realization of this idea can be achieved with
a complex scalar field $\Phi$ acquiring different complex vacuum
expectation values:
\be
\langle \Phi \rangle = \pm f
\label{vac-phi}
\ee
This can be done if the potential of $\Phi$ has 
two degenerate minima at $\Phi=\pm f$. It resembles more complicated
potential of the electroweak Higgs field but instead of continuous
symmetry this potential has only a discrete symmetry with respect to
the transformation $\Phi \leftrightarrow -\Phi$.

Such a mechanism of CP violation is locally indistinguishable from
the explicit one but globally it leads to charged symmetric
universe with domains of opposite but equal by magnitude baryon 
asymmetry. As we mentioned above in a globally baryo-symmetric universe 
matter-antimatter domains should be very large  with $l_B\sim $ Gpc to
avoid too strong gamma ray background. In the case of spontaneous CP
violation there is one more reason to make the domain size very big,
even larger than the present day horizon. If so, our chances to observe 
cosmic antimatter are negligible at least during the next several 
billion years. The condition of a huge domain size is related to a
well known problem of cosmological domain walls~\cite{zko},
which arises if a discrete symmetry is spontaneously broken.
The energy of the wall separating two different vacua with opposite
signs of CP-violation is so large that even a single wall per horizon
volume would destroy the observed isotropy of the universe. 
This domain wall problem either demands $ l_B \gg $ Gpc or some
non-trivial mechanism of wall destruction after baryogenesis.

\subsection{Stochastic or dynamical CP violation 
\label{ss-stoch}~\cite{ad-baryo-rev}}

This mechanism is somewhat similar to spontaneous 
CP breaking but operates only in the early universe and does not suffer 
from domain wall problem. It could be especially good for creation
of subdominant antimatter domains or 
compact objects in our matter dominated
universe~\cite{ad-anti-m}. Similar versions of dynamical
CP-violation have been studied recently in papers~\cite{dyn-cp}.

Let us assume that there exists a complex scalar field $\chi$ by 
some reason displaced from its equilibrium  point, e.g. by quantum 
fluctuations at inflation. It is known that
a massless or light, $m<H$,
scalar field is infrared unstable at De Sitter stage and its average
value rises with time as~\cite{infrared}:
\be
 \chi^2 \sim H^3 t.
\label{chi-2}
\ee
(see however ref.~\cite{ad-dp} where it was argued that the coefficient in 
eq. (\ref{chi-2}) has an opposite sign but the same absolute value).

When inflation is over, $\chi$ starts to relax down to zero.
If this relaxation process is not too fast so that $\chi \neq 0$
during baryogenesis, $\chi$ would act as a CP-violating agent. 
Its complex
amplitude forces time dependent CP-violation into action. In the old
universe $\chi$ either vanishes and all CP-violating effects 
associated with $\chi$ vanishes as well, or $\chi$ may remain
nonzero and create the usual explicit CP-violation.
As we see in what follows with such a mechanism of CP-violation
an inhomogeneous $\beta(x)$ could be easily generated. In particular,
domains with $\beta <0$, that is antimatter domains, might be created. 
They could be not too far from us and in sufficiently small amount
to satisfy the existing constraints. Moreover, if antimatter is mostly
confined inside compact stellar type objects, the upper bound on their 
cosmological mass density could be much less restrictive.

Dynamical C(CP)-violation could create a very interesting/unusual 
pattern of the cosmological matter-antimatter distribution. For 
example, if a complex scalar field, $\phi$, oscillated around its 
equilibrium point at $\phi=0$ during baryogenesis then it would
contribute to CP-odd amplitude 
in addition to the standard explicit CP-violation term. If the
potential of $\phi$ is not strictly harmonic (quadratic) one, then
the baryon asymmetry would have a spatially periodic 
distribution~\cite{mc-ad}. Depending upon the relative magnitude
of explicit and dynamical CP-odd terms there would be either 
periodic layers of matter and antimatter in the universe separated
by voids with a small baryon or antibaryon number density, or
some periodic modulation of homogeneous baryonic background.
The voids between the matter and antimatter domains may help to
loosen the bound of ref.~\cite{cdg} that the nearest domain of
antimatter in baryo-symmetric universe should be at the distance 
above 1 Gpc. On the other hand, such fluctuations in 
baryon/antibaryon number density would lead to large isocurvature 
fluctuations which contradict the observed spectrum of angular
fluctuations of CMBR at the scales down to tens of Mpc. At smaller
scales the bounds from CMBR are weak because of the diffusion
(Silk) damping~\cite{silk} but such scales are restricted,
though not so strongly, by
the observations of the large scale structure of the universe.

Some more examples with dynamical CP-violation are discussed at the 
final part of these lectures (see below, secs. \ref{ss-spont},
\ref{ss-susy}).

\section{Electroweak baryogenesis in the minimal standard 
model~\cite{krs} \label{s-ew}}


In the standard electroweak (EW) theory
all the necessary ingredients for baryogenesis are present:\\
1. CP is known to be broken (but very weakly, see sec.~\ref{ss-explct}).\\
2. Baryonic charge is non-conserved because of chiral 
anomaly~\cite{thooft}. At
zero T baryon nonconservation is exponentially suppressed,
as $ \exp (-2\pi/\alpha)$, because baryon nonconservation proceeds
through barrier penetration between different vacua.
At high T it might be possible to go over the barrier,
but abundant formation of specific classical field
configuration, sphalerons, is necessary. Classical field
configurations are those whose size is much larger than their
Compton wave length and quantum production of such objects, which
are coherent at very large distances, is not well understood,
see e.g. discussion in ref.~\cite{ad-taup}.\\
3. Thermal equilibrium is absent during the phase transition between
the phases with unbroken EW symmetry at high $T$ and broken one 
at low $T$ if the phase transition
is first order. However, a heavy Higgs renders the 
first order phase transition improbable.
Another possible source of deviation from equilibrium due to massive 
particles is weak: $\sim m_{EW}/m_{Pl}\sim 10^{-16}$. However it
could be large in TeV-scale gravity.

CP-odd amplitude in MSM is estimated above, eq. (\ref{a-odd}).
At high temperatures near EW phase transition the characteristic 
mass should be of the order of the EW energy scale,
$ M\sim 100$ GeV, and thus
\be
A_- \approx 5\cdot 10^{-19}.
\label{a-high-T}
\ee
Such a smallness of CP violation at high temperatures makes
hopeless generation of the observed baryon asymmetry in the
frameworks of MSM. 

The result (\ref{a-high-T}) is rather surprising because at 
${ T=0}$ the effects of CP are non-negligible. Naively one might
expect that CP-odd effects in a decay of a heavy quark should
not be much different at high and low $T$ because the propagators,
which determine the normalizing mass $M$
in the amplitude of decay ``know'' the mass of the decaying particle
and not its energy. However, this is not the case because of the
temperature corrections to the mass, or better to say to the dispersion
relation, $E=E(p,T)$. The largest corrections come from
QCD because of the largest coupling constant. This effect leads to
a large denominators of the propagators which enters the CP-odd
amplitude and thus leads to a large $M\sim 100$ GeV, while
in the numerators there are still ``Higgs masses'' or, better to say,
small Yukawa coupling constants.
Attempts to modify quark dispersion relation at high T~\cite{far-shap}
were proved to be unsuccessful~\cite{cern}.

Surprisingly a much larger amplitude of CP-violation found in
ref.~\cite{ibk-cp} at zero temperature gives the same result
as eq. (\ref{a-high-T}) at high $T$.

Thus it seems that MSM is unable to explain the observed cosmological 
baryon asymmetry and physics beyond the standard model is necessary.
The standard CP-odd effects may be strongly amplified if Yukawa
coupling constants change with time in such a way that they are
much larger at high T. Such a mechanism is suggested in
ref.~\cite{nir}. The first order phase transition,
which is necessary to break thermal equilibrium, still remains
problematic but with a low value of the Planck mass at EW scale
one can manage with second order phase transition as well.

\section{Baryogenesis through heavy particle decays}

The original idea of this type of baryogenesis stems from the
pioneering works~\cite{ads,kuzmin}. A large list of literature
can be found e.g. in the reviews~\cite{ad-baryo-rev,ad-yz}.
From particle physics perspective, two most conservative 
models of this kind are GUT baryogenesis and baryo-thru-leptogenesis.

Before going to more detailed discussion of these two models
we describe some general features of generation of cosmological
charge asymmetry by decay of heavy particles. Let us consider 
the decays of a heavy boson $X$ (it can be e.g. gauge or Higgs boson
of grand unification) into two decay channels with different
baryon numbers:
\be
X \rightarrow qq, \,\,\, X \rightarrow  q\bar l,\nonumber\\
\bar X \rightarrow \bar q \bar q, \,\,\,
\bar X \rightarrow \bar q  l \, .
\label{x-decays}
\ee
The widths of the charge conjugated channels would be
different only in higher orders of perturbation theory
when rescattering {with baryonic charge 
non-conservation} in the final state is taken into
account (see conclusion to sec.~\ref{s-equil}):
\be
\Gamma_{X\rightarrow qq} = (1+\Delta_q) \Gamma_q, \,\,\,
\Gamma_{X\rightarrow  q \bar l} = (1-\Delta_l) \Gamma_l, \nonumber \\
\Gamma_{\bar X \rightarrow \bar q \bar q } = (1-\Delta_q) \Gamma_q, \,\,
\Gamma_{\bar X \rightarrow \bar q l} = (1+\Delta_l) \Gamma_l.
\label{delta-gammas}
\ee
Here $\Delta_{q,l}$ are non-zero due to CP violation and if other channels
of $X$-decays are negligible, then 
$\Gamma_q\Delta_{q} =\Gamma_l\Delta_{l}$ to ensure the
equality of the total decay widths of $X$ and $\bar X$ which follows from
CPT invariance.

As we have already mentioned, 
particles and antiparticles can have 
different decay rates into charge 
conjugated channels, if both C and CP are broken. 
If only C is broken, but {CP is OK,} then partial widths, summed over
spins, are the same:
\be
\Gamma \left( X\rightarrow f, \sigma \right) =
\left( \bar X\rightarrow \bar f, -\sigma \right) 
\label{sum-spins}
\ee
Thus to break equality of the partial decay widths
both C and CP should be broken.

If the initial state is populated only by $X$ and $\bar X$ bosons in equal
amount and if only their decays are essential and all other processes 
can be neglected, the baryon asymmetry generated by these decays
would be evidently proportional to
$ (2/3)(2\Delta_q -\Delta_l)$.

Such an initial state could be realized if by some reason inflaton 
quickly decayed into $X$ and $\bar X$ and not into anything else. In 
reality,
more probable is high temperature cosmic plasma which is rather close
to thermal equilibrium. At this stage we may ask the question
how baryon asymmetry which is generated in the decays of $X$ and $\bar X$
would be equilibrated down to zero? What processes are responsible for that?
A usual (and incorrect) answer is that the inverse decay, $qq \rar X$, and 
similar ones do the job. 
However using CPT, one finds:
\be
\Gamma_{\bar q \bar q \rightarrow \bar X } = (1+\Delta_q)\Gamma_q,
\,\,
\Gamma_{\bar q l \rightarrow \bar X } = (1-\Delta_l)\Gamma_l,
\nonumber \\
\Gamma_{ q  q \rightarrow  X } = (1-\Delta_q)\Gamma_q,
\,\,
\Gamma_{ q \bar l \rightarrow  X } = (1+\Delta_l)\Gamma_l.
\label{decays-cpt}
\ee
Thus direct and inverse decays produce the {\it same} sign of baryon
asymmetry! It can be shown that destruction of the baryon asymmetry is
achieved by baryon non-conserving $qq$ and/or $ql$ scattering with 
exchange of $X$ and $\bar X$ bosons.

Let us comment a little more on the necessity of rescattering for
the generation of the baryon asymmetry. From the unitarity 
condition (\ref{im-T}) follows that if only two reaction 
channels $i$ and $f$ are open then:
\be
2 {\cal I}m T_{ii}[\lambda] = -\int d\tau_i |T_{if}|^2 -
\int d\tau_f |T_{ff}|^2
\label{ii}
\ee
By CPT transformation:
\be
T_{ii}[\lambda] = T_{\bar i \bar i}[-\lambda]
\label{t-ii}
\ee
and after summing over polarization we find 
$ \Gamma_{if}=\Gamma_{\bar i \bar f}$. 
Hence to destroy the equality of partial widths of charge conjugated 
processes,  $\Gamma_{if} = \bar \Gamma_{\bar i \bar f}$,
at least three reaction channels must be open.
\be
i \lrar f,\,\,\, i \lrar k,\,\,\, k \lrar f.
\label{ifk}
\ee

Let us discuss now some concrete examples.

Grand unified theories (GUT) naturally violate baryonic number
conservation because quarks and leptons are put into the same
multiplet of the symmetry group. The mass of the gauge bosons of grand
unification, $m_X\sim 10^{16}$ GeV, is high enough to ensure 
sufficiently large deviation from equilibrium, see eq. (\ref{delta-f}).
CP-violation can be easily unsuppressed at grand unification scale,
$T\sim m_X$. So far so good, but such high temperatures may never 
be accessible in the universe after inflation but even if 
the universe was hot enough with $T>m_x$
or X-bosons could be produced out of equilibrium,
one should take care of overabundant production of 
gravitinos~\cite{gravitino}. The latter of course could be dangerous 
only if supergravity is realized.

At the present time baryo-thru-leptogenesis~\cite{fuk-yan} is probably 
the most popular scenario of creation of baryo-asymmetric universe.
The process proceeds in two steps.
First, lepton asymmetry is generated in decays of a heavy Majorana
neutrino, $N$, which was postulated for realization of the seesaw 
mechanism. Evidently the decays of $N$ do not conserve leptonic charge.
At the next step, during the electroweak stage, the lepton asymmetry is 
transformed into baryon asymmetry by C and CP conserving sphaleron 
processes in thermal equilibrium. Sphalerons do not conserve 
baryonic, $B$, and leptonic, $L$, charges individually but they conserve
$(B-L)$. Thus initial $L$ would be redistributed in equilibrium in
almost equal shares between $B$ and $L$. For reviews of this scenario
see refs.~\cite{lept-rev}.

This mechanism looks very attractive. 
Leptonic and baryonic charges are naturally nonconserved.
Heavy particles (Majorana neutrinos) to break thermal equilibrium 
are present. Three CP-odd phases of order unity might be there.
However, the magnitude of the asymmetry $\beta$ is just of the right 
size if the situation is most favorable. Any deviation from the most
favorable case would destroy the successful prediction of the model.

As we mentioned above, the deviation from thermal equilibrium is given
by eq. (\ref{delta-f}). The magnitude of the lepton asymmetry 
can be estimated as
\be
\beta_L \sim \frac{\delta f}{f}\,\frac{\Delta \Gamma}{\Gamma}
\sim \frac{m}{m_{Pl}}
\label{beta-l}
\ee
where $\Delta \Gamma$ is the difference of the $L$-violating decay
widths of charge conjugated channels.
Some small numerical coefficients, which should be
present in eq. (\ref{beta-l}), would make the result even smaller.
Subsequent entropy dilution due to annihilation of massive particles
in the course of cosmological expansion and cooling down could diminish
this result by about 1/10-1/100.  
Taken all together,
for successful lepto/baryo-genesis the mass of the decaying 
heavy Majorana lepton 
should be noticeably larger than ${ 10^{10}}$ GeV
(or ${ m_{Pl} \ll 10^{19}}$ GeV at high energies).
An additional entropy generation by possible, especially first order,
phase transitions would make the final baryon asymmetry considerably 
smaller.

An interesting suggestion has been put forward in ref.~\cite{lind}. 
It was assumed that there exist new heavy scalars decaying into 
right-handed neutrinos and left-handed leptons (charged or neutral).
By assumption, leptonic charge is conserved in these decays and the
total cosmological lepton asymmetry remains zero till electroweak
sphalerons start to operate. Sphalerons would distribute leptonic charge
density of right-handed leptons into almost equal shares between
leptonic and baryonic asymmetries as is described above. On the
other hand lepton asymmetry of right-handed, sterile neutrinos
would remain untouched and unnoticed. In the concrete realization of the
model the ``natural'' value of the neutrino mass due to loop diagram with
an exchange of the hypothetical heavy scalars and the charged tau-lepton
should lead to too large value of neutrino mass, so some fine-tuning
at the level of 0.01-0.001 is necessary. Possibly in other realization of
this idea this shortcoming could be avoided.

\section{Some more models of baryogenesis}

There are plenty of more speculative scenarios of baryogenesis, 
than the ``conservative'' ones presented above, but who knows, 
maybe one of these speculative scenarios happens to be true.
It is rather difficult to exclude or confirm one or other scenario
because they all need to explain only one number, $\beta$. However,
if $\beta$ is not just a number but a function of space points,
which creates some specific isocurvature perturbations and even
some domains or objects of antimatter (if $\beta$ somewhere becomes 
negative), the odds to discover the truth become much higher. Another
good chance to check the baryogenesis scenario arises if it predicts
a certain form of dark matter which is correlated with generation of 
baryon asymmetry. All scenarios presented above do not present
any natural explanation of
the close values of the energy density of baryons and that of dark 
matter:
\be
\rho_B / \rho_{DM} \approx 0.2
\label{r-dm}
\ee
Most probably this coincidence is not accidental - these two numbers may
easily differ by many orders of magnitude. It is desirable that a 
realistic scenario of baryogenesis is able not only to present the
right value of $\beta$ but also to explain the
magnitude of the ratio (\ref{r-dm}). However, maybe our  
knowledge of particle physics is not good enough for that.

Below we consider scenarios of baryogenesis which may
have, though not necessarily, some of the features described above.

\subsection{Baryogenesis through evaporation of primordial black holes }

This model  does not demand B-nonconservation at
particle physics level. If a conserved charge does not create
any long-range field, as e.g. electric charge does, then such
a charge could disappear inside a black hole (BH) without trace. If by
some reason black holes prefer to capture predominantly
antibaryons then in external space an excess of baryons over 
antibaryons would be generated. These black holes could be
heavy enough to survive till our time, $t_u \approx 10$ Gyr.
Another possibility discussed
in refs.~\cite{hawk-bs,zel-bs} is that the process of BH
evaporation~\cite{evap}
could be baryo-asymmetric and small black holes in the process of
their evaporation would enrich the universe with baryons. Such
black holes could either disappear completely or evolve down to
stable(?) Planck mass remnants but in all these cases the universe
outside black holes would have a non-vanishing baryon asymmetry and
the equal amount of antibaryonic charge would be buried inside
black holes which either completely disappeared or survived and  
became cosmological dark matter. 

At first sight thermal evaporation of black holes
cannot create any charge asymmetry by the same reason as
charge asymmetry is not generated in thermal equilibrium.
However the spectrum of particles radiated by black holes
is not black but gray due to propagation
of the produced particles in the gravitational field of black 
holes~\cite{page}.
Moreover, interactions among the produced particles are
also essential.
These two facts allow black holes to create an excess
of matter over antimatter in external space. 
As a possible ``realistic'' model 
let us consider the following~\cite{zel-bs,ad-bh}. 
Let us assume that there exists a heavy $A$-meson which may
decay into two charge conjugated channels with unequal
probabilities (due to C and CP violation):
\be
A\rar H+\bar L \,\,\,{\rm { and}}\,\,\, A\rar \bar H + L
\label{A-decays}
\ee
where $H$ and $L$ are respectively heavy and light baryons, e.g.
$t$ and $u$ quarks.

If the temperature of a black hole is larger than or comparable to
the mass of A-boson the latter
would be abundantly produced at the horizon and decay
while propagating in the gravitational field of BH. There is a non-zero 
probability of back capture of the decay products by BH and evidently
the back-capture of the heavy baryons, $H$ and $\bar H$, is larger than
that of the light ones, $L$ and $\bar L$. 
As a result a net baryon asymmetry could be created outside black holes.
According to ref.~\cite{ad-bh} the baryon asymmetry 
may have the proper magnitude compatible with observations.

If at the moment of BH formation, which presumably took place at radiation
dominated (RD) stage, primordial black holes contributed
a very small fraction, $\epsilon$, to the total cosmological energy
density, then at red-shift $z=1/\epsilon$ after formation, black holes
would dominate cosmological energy density if their life time happened
to be larger than the time interval necessary for this red-shift,
$t_{MD} = t_{in} /{\epsilon}^2$. Note, that 
the black hole formation might take 
place at some earlier matter dominated (MD) stage which turned into RD stage
later. In this case the estimated time would be 
somewhat different but the 
scenario would still be viable. Black hole evaporation at 
$\tau_{BH}> t_{MD}$ would recreate radiation dominated universe but
now with non-vanishing baryon asymmetry. Here $\tau_{BH}$ is the 
life-time of BH with respect to evaporation.

Now for the convenience of listeners
I present some order of magnitude expressions for the 
quantities describing BH evaporation. Numerical coefficients
of order unity (or sometimes $4\pi$)
are omitted. Precise expressions can be found 
in any modern text-book on BH physics, e.g.~\cite{fn}. For a 
Schwarzschild black hole the essential dimensional parameter is its
gravitational radius:
\be
r_g = M_{BH}/m_{Pl}^2
\label{rg}
\ee
The black hole temperature, just on dimensional grounds, is the
inverse gravitational radius:
\be
T_{BH}\sim 1/r_g \sim m_{Pl}^2/M_{BH}
\label{t-bh}
\ee
The luminosity can be easily estimated integrating black body radiation
emitted by the object of size $r_g$:
\be
L_{BH} \sim T^4 r_g^2 \sim m_{Pl}^4/M^2_{BH}
\label{l-bh}
\ee
Knowing the BH mass and luminosity it is straightforward to
estimate its life-time:
\be
\tau_{BH} \sim M_{BH}^3/m_{Pl}^4
\label{tau-bh}
\ee
Black holes with the mass of the order of ${ M_{BH} = 10^{15}}$g
would have radius approximately $10^{-13}$ cm and temperature about
100 MeV. They could survive to the present time,  
$ \tau_{BH} \approx t_U \approx 10$ Gyr.

According to the calculations of ref.~\cite{ad-bh} the mass of
heavy decaying particles should be in the interval
{${ m\sim 10^{10}-10^{6}}$ GeV} to create the observed
cosmological baryon asymmetry.

Let us consider the following example. Assume that primordial
BH were created when temperature of the universe was about 
$10^{14}$ GeV. It corresponds to the universe age 
$t_U \approx 10^{-34}$ sec.
The mass inside horizon at that moment was
\be
M_h \approx 10^{38}\,{\rm g}\,(t/{\rm sec}) \approx10^4\,\,{\rm g}. 
\label{m-hor}
\ee
Black holes with such masses might be in principle created. Their temperature 
and life time would be respectively:
\be
T_{BH} = 10^{10} \,{\rm GeV}
\label{temp-bh}
\\
\tau_{BH} \sim 10^{-16}\,\,{\rm sec,}
\label{time-bh}
\ee
During this time the universe would cool down to
{${ T \sim 10^5}$ GeV} and the red-shift from the moment
when horizon mass was equal to ${ M_{BH}}$, would be about
${z \sim 10^{10}}$.

If the black hole production efficiency
was such that only in 1 per $10^{10}$
horizon volume a black hole was created, then their
mass fraction at production was $ 10^{-10}$ and at the moment
of their evaporation they would dominate cosmological energy density and
{could create the observed baryon asymmetry.} As we mentioned above the
Planck mass remnants of such primordial BH could be cosmological 
dark matter, see e.g.~\cite{ad-nas} .

If baryon asymmetry could be generated via evaporation of a classical
(in contrast to quantum) black holes, then it is natural to expect 
that baryonic charge might be non-conserved also in decay of small
quantum black holes. Expressed in other words it means that
gravity breaks all global symmetries and at 
the Planck scale the effects should
be unsuppressed. First this observation was done in ref.~\cite{zel-decay}
where it was stated that proton must decay due to transformation 
into a virtual black hole which subsequently decays 
into, say, positron and meson. We can
very crudely estimate the proton life-time as follows (it differs
from the original Zeldovich estimate into much larger direction).
The probability that one quark would collapse into a virtual black hole
is proportional to the volume of such
black hole, i.e. to $r_g^3 \sim m_q^3/m_{Pl}^6$. However, because
of charge and color conservation this black hole must be stable.
Thus to force proton to decay, two quarks in proton should enter inside 
horizon and to form ``two-quark'' black hole
which may decay into antiquark and lepton. The probability
of such process is further suppressed by a power of $(m_q/m_{Pl})^n$.
It is unclear what should be substituted here instead of effective 
quark mass, its bare (current) mass, $m_q \approx 5$ MeV, or its 
constituent mass, $m_q \approx 300 $ MeV. Possibly in the first factor
the bare mass (mass at small distances) should be substituted, while the
second factor might contain the constituent mass.
Even with the constituent mass the proton life-time
with respect to such gravitational decay would be huge:
\be
\tau (p\rar e^+ \pi^0) 
\sim \left(r_g^3 m_p^4\right)^{-1} = m_{Pl}^{6+n}/m_{p}^{7+n}
\approx 10^{90}\,{\rm { sec}} ( m_{Pl}/m_{p})^n
\label{p-grav}
\ee
What happens, however, if TeV gravity is realized, i.e. $ m_{Pl}\sim$ TeV?
If in the estimate above one should substitute small bare quark mass then
the proton life-time even for $n=2$ would be $\sim 10^{44}$ sec which
is completely safe. However, heavy quarks, e.g. $t$-quark may 
noticeably decay into $2\bar q+l$. A discussion of heavy particle
decay and a list of literature can be found in the lecture~\cite{ad-erice}.

\subsection{Baryon asymmetry with conserved baryonic charge}

Though it was claimed in the previous section that baryonic charge
is conserved in the process of large (classical) black hole evaporation,
strictly speaking, 
it was not so. If the final result of black hole evaporation
would be a stable Planck mass remnant, one might say that the 
compensating amount of (anti)baryonic charge is stored in these
remnants. Probably 
in quantum world these particles
should be indistinguishable, despite different 
baryonic charges, and we must admit that
at quantum level baryonic charge is not conserved, though it is formally
conserved in particle physics Lagrangian. 
Moreover, proton decay through formation of 
a  virtual black hole surely demonstrates the non-conservation of baryons.

Now we consider a model of baryogenesis where baryons are really
conserved~\cite{conserv-b}, see also~\cite{ad-baryo-rev}.  
This model is somewhat similar to the
one discussed above but without black holes. Their role now plays
a sterile (or weakly interacting with our world) new baryon $Q$.
Baryon asymmetry can be e.g. generated in decays of some heavy boson:
\be
A \rar q + \bar Q,\,\,\,{\rm { and}}\,\,\,
\bar A \rar Q + \bar q
\label{aqQ}
\ee
with different partial widths of charge conjugated channels. It
would lead to equal but opposite signs of 
baryon asymmetries in ours and in (hidden) Q sector. A nice 
feature of this model is that
{Q-baryons could make dark matter} if their mass is 5 times larger 
than ${ m_p}$. So we have
a unique mechanism of baryogenesis and DM creation. Unfortunately 
there is no indication from particle physics to such a new baryon.

The scenario reminds generation of large lepton asymmetry in
the sector of active neutrinos through resonance oscillations
between active and hypothetical sterile neutrinos~\cite{nua-nus}.
Leptonic charge is assumed to be conserved but the probability of
transformation of active neutrinos into sterile partners is 
different from the similar process for antiparticles. This process goes
without an explicit CP-violation in neutrino mass matrix. The
necessary breaking of symmetry between particles and antiparticles is
induced by the charge asymmetry of the cosmic plasma and corresponding 
charge asymmetric terms in the neutrino refraction index. 

The similar scenario may operate at high temperatures if there exist
heavy neutrinos, one of which being sterile with respect to our world
but unstable, decaying into sterile sector. Since by assumption the
cosmological plasma at this stage was charge symmetric, one needs
CP-violation. The latter may be introduced directly into neutrino
mass matrix. As we have seen above there is plenty of space for
CP-odd phases, see the end of sec.~\ref{ss-explct}.
Now lepton asymmetry in active neutrino sector could be generated
by heavy neutrino oscillations. Due to conservation of the total
leptonic charge equal and opposite asymmetry would be created in the
sector of heavy sterile neutrinos. Later leptonic charge of active 
neutrinos would be transformed into baryonic charge by electro-weak
processes as it happens in baryo-thru-leptogenesis scenario,
sec.~\ref{s-ew}. This is another example of generation of baryon
asymmetry with conserved charges.

A few months after this School a paper~\cite{bauer} has appeared with
somewhat similar idea that the difference of baryonic and leptonic charges,
$(B-L)$, is globally conserved, while the observed charge asymmetry in
the standard particle sector is compensated by $(B-L)$ stored in dark
energy sector.

\subsection{ Spontaneous baryogenesis \label{ss-spont}}

If a global symmetry, $ U(1)$, related to baryonic or some other
charge which is not orthogonal to baryonic one, is spontaneously
broken, the spontaneous baryogenesis scenario~\cite{spont-bs} could
be realized. As usually spontaneous symmetry breaking is described
by the Higgs mechanism. A scalar field $\phi$ with nonzero baryonic 
charge acquires a vacuum expectation value because its potential
energy reaches minimum at non-zero $\phi$. 
\be
U(\phi) = \lambda (|\phi|^2 - \eta^2)^2 ,
\label{u-of-phi}
\ee
Though the potential is invariant with respect to $U(1)$ phase rotation,
$\phi\rar \exp (i\theta) \phi$, an accidentally chosen ground state,
$\langle \phi\rangle \exp (i\theta_0) \eta$, is not. 
Spontaneous breaking of a
global symmetry leads to appearance of a massless scalar field, 
Goldstone boson, proportional to the phase of $\phi$
\be
\phi = \left[\eta+\zeta (x)\right] \exp [i\theta(x)]
\label{phi-theta}
\ee

The Lagrangian of the Goldstone field, $\theta (x)$, has the form
\be
{\cal L} =\eta^2 (\partial \theta )^2 + 
\partial_\mu \theta j^B_\mu
- V(\theta) + i\bar Q \gamma_\mu \partial_\mu Q + 
i\bar L \gamma_\mu \partial_\mu L +
(g \eta \bar Q L + h.c.).
\label{l-theta}
\ee
The potential $V(\theta)$ may be nonzero due to
some other physical effects. If $V''\neq 0$ the field
$\theta$ would be massive, but usually light. In this case 
it is called pseudogoldstone boson.

In the homogeneous situation the second term in eq. (\ref{l-theta})
looks like chemical potential,
$\dot\theta n_N$, but in
reality it is not, because the coupling is derivative
and the corresponding term in the Lagrangian (\ref{l-theta}),
$\partial_\mu \theta j^B_\mu$,
does not coincide
with that in the Hamiltonian~\cite{ad-kf}.

If ${ V(\theta) =0}$, i.e. purely Goldstone case, we can
integrate the equation of motion:
\be
2 \eta^2 \partial^2 \theta = -\partial_\mu j_\mu ^B
\label{eqn-theta}
\ee
and obtain:
\be
\Delta n_B = - \eta^2 \Delta \dot \theta,
\label{delta-b-gold}
\ee
where $\Delta$ means the difference between the running values and
the initial ones. Thus, non-zero baryon asymmetry may exist
in thermal equilibrium and without
explicit CP-violation. The latter is created by initial
$ \dot\theta\neq 0 $. This is an example of dynamical CP
violation in cosmology discussed in sec. \ref{ss-stoch}.

In realistic situation $ \dot\theta$ is small, because kinetic
energy is normally red-shifted away in the course of expansion,
and hence the pseudogoldstone case could be more efficient for creation
of the baryon asymmetry.
The equation of motion in the pseudogoldstone case has the form:
\be
\eta^2 \ddot \theta +3H\dot \theta +  V' (\theta) =
 \partial_\mu j_\mu^B.
\label{eq-pseudo}
\ee
where the second term proportional to the Hubble parameter $H$
appears because of the cosmological expansion, this is the so
called Hubble friction term, 
${ j_\mu^B = \bar\psi \gamma_\mu \psi}$ is the baryonic currents 
of fermions (quarks) and the potential is expanded near
minimum (which is taken to be at $\theta =\pi$) as:
\be
V(\theta) \approx -1+m^2_\theta \eta^2 (\theta-\pi)^2 /2
\label{v-theta}
\ee
where $m_\theta$ is the mass of the pseudogoldstone field.

Initially, at inflation, $\theta$ should be uniformly distributed 
in the interval $ [0,2\pi]$, so at the end of inflation it would
typically have the value of order unity. When the universe
expansion slowed down so that $H$ becomes smaller than the mass
of $\theta$, the field started to evolve down to the equilibrium
point of the potential oscillating around it. 
The value of the baryon asymmetry which
was generated on the way is not so easy to calculate as 
in the purely Goldstone case. To this end the equation of motion
of fermions has to be solved.
The equation for the quantum operators of the
baryonic Dirac field reads:
\be
\left(i\partial+m  \right)\psi = -g\eta l+
(\partial_\mu\theta) \gamma_\mu \psi
\label{dirac}
\ee

One can find the solution of this equation 
in one-loop approximation for $\psi$ in external classical
field $\theta$ and substitute ${ \bar\psi \psi = F (\theta)}$ into
equation of motion for $\theta$ (\ref{eq-pseudo}).
The solution oscillates with alternating baryonic number.
Because of these oscillations the net result for the 
baryon asymmetry is not linear in $\theta$  but cubic~\cite{dfrs}:
\be
n_B \sim \eta^2 \Gamma_{\Delta B} (\Delta \theta)^3 .
\label{b-pseudo}
\ee
where $\Gamma_{\Delta B}$ is the decay rate of $\theta$ into
baryons and $\Delta \theta$ is the difference between initial
value of $\theta$ and the equilibrium one at the minimum of the
potential $V(\theta)$.

\subsection{Baryogenesis from super-partner baryonic 
condensate \label{ss-susy}}

As is well known, supersymmetry (SUSY) predicts existence of scalar
superpartners of baryons which have non-zero baryon number.
Self-potential of such scalars can typically have flat directions
along which potential energy does not change. Moreover, baryonic
charge is naturally non-conserved. Some more detail about SUSY are
presented in the lecture by A. Masiero at this School.

As a toy model possessing these properties we can take the scalar
partner of baryons, $\chi$ with the potential:
\be
U_\lambda(\chi) = \lambda |\chi|^4 \left( 1-\cos 4\theta \right),
\label{u-of-chi}
\ee
where ${ \chi = |\chi| \exp (i\theta)}$.
There are four flat directions in this potential along 
$\cos 4\theta = 1$. The potential breaks symmetry with respect
to phase rotation $\chi \rar \chi \exp (i\alpha)$. This leads
to nonconservation of baryonic charge of $\chi$.
Due to infrared instability of massless fields in De Sitter space-time,
see eq. (\ref{chi-2}), such bosons may condense along flat 
directions of the potential. To be more accurate, only colorless and 
electrically neutral combination of the fields may condense.
When these flat directions acquired a non-zero curvature (mass) and 
the universe expansion rate becomes smaller than the mass, $H<m_\chi$,
the field would evolve down to equilibrium. During this relaxation
down to $\chi =0$ the field $\chi$ decays into quarks, most probably
with conserved $B$, and released the baryonic charge stored in 
the condensate into the baryonic charge of quarks. This is an essence
of the Affleck and Dine scenario of baryogenesis~\cite{af-di}.

In addition to the quartic potential (\ref{u-of-chi}) the
mass term can be added:
\be
U_m(\chi) = m^2 |\chi|^4 \left[ 1-\cos \left(2\theta+2\alpha\right) \right],
\label{u-m}
\ee
Here $\alpha$ is some arbitrary phase.
If $\alpha \neq 0$, C and CP are explicitly broken.
 
``Initially'' (at inflation) $\chi$ is away from the origin and when
inflation is over, $\chi$ starts to evolve down to the equilibrium point, 
$\chi =0$, according to its equation of motion which for homogeneous 
$\chi$ coincides with equation of motion of point-like particle in
Newtonian mechanics:
\be
\ddot \chi +3H\dot \chi +U' (\chi) = 0.
\label{eqn-chi}
\ee
Baryonic charge of $\chi$,
\be
B_\chi =\dot\theta |\chi|^2
\label{b-chi}
\ee
is analogous to mechanical angular momentum. When $\chi$ decays, its
baryonic charge is transferred to that of quarks in B-conserving process.
Thus we can easily visualize the process without explicit solution of
the equations of motion.

The B-charge of $ \chi$ is accumulated
in its ``rotational'' motion,
induced by quantum fluctuations in the orthogonal to valley direction.
The space average value of the baryonic charge is evidently zero
and as a result a globally charge symmetric universe would be
created.
The  size of domains with definite sign of the baryonic charge
density, $l_B$, is determined by the size of the region
with a definite sign of $\dot\theta$. Normally the size of the region
with definite $\dot\theta$ is microscopic and this leads to a very 
small $ l_B $. However, if the Hubble parameter at inflation happened
to be larger than the second derivative of the potential $U_\lambda$
in the direction orthogonal to the valley, the field motion in 
orthogonal direction would be frozen during exponential expansion and 
the size of the domains with a fixed value of $B$ may be large
enough.

Situation may be different if ${ m\neq 0}$.  In this case initial
angular momentum or, what is the same, baryonic charge of $\chi$
could be zero but the rotational motion (or baryonic charge)
may  be created by a different direction of the valley at low $\chi$.
At large $\chi$ the direction of the valley is determined
by $U_\lambda (\chi)$, eq. (\ref{u-of-chi}), while at small $\chi$
the quadratic part (\ref{u-m}) dominates.

If the
CP-odd phase in eq. (\ref{u-m}) is zero, $\alpha =0$, but the flat 
direction of $U_\lambda$, along which $\chi$ condensed, is
orthogonal to flat directions of $U_m$, the field $\chi$ would
rotate near origin
with 50\% probability clock-wise or anti-clockwise creating
baryonic or antibaryonic universe. If inflation helped, such regions 
could be sufficiently big. This is an example of baryogenesis without
an explicit C(CP)-violation and without domain wall problem.

If the CP-odd phase, $\alpha$, is small but non-vanishing, the rotation
of $\chi$ when it approached the $m$-valley would proceed with
different probabilities in different directions. Hence
both baryonic and antibaryonic regions are possible with a dominance 
of one of them. Matter and antimatter domain may exist but globally 
${ B\neq 0}$~\cite{ad-anti-m,ad-js}. 
A different scenario of formation of cosmic domains of
antimatter can be found in the paper~\cite{khlop} (and references
therein).

A very interesting picture appears if the field $\chi$ is coupled
to the inflaton with the general renormalizable coupling~\cite{ad-js}:
\be
\lambda |\chi|^2 \left( \Phi - \Phi_1 \right)^2
\label{chi-infl}
\ee
In this case the ``gates'' to the 
the valley may be open only for a short time when the inflaton
field $\Phi$ was close to $\Phi_1$. Thus the probability of penetration
to the valley would be small and $\chi$ would acquire
a large baryonic charge condensate, giving $\beta \sim 1$
only in a tiny fraction of space.
The bulk of space would have the  
normal homogeneous baryon asymmetry
${ \beta = 6\cdot 10^{-10}}$, which could be created by one of the
standard mechanisms described above,
while hi-B regions would be rare. Depending upon the concrete model, the
hi-B regions may be symmetric with respect to baryons and antibaryons
or dominated by one of them.

The mass distribution of the hi-B regions was 
calculated in ref.~\cite{ad-js}. It depends upon two unknown constants
$C_0$ and $C_1$ 
and has a simple log-normal form:
\be
\frac{dN}{dM} = C_0 \exp\left[ -C_1 \ln^2 \left(M/M_0\right)\right].
\label{dn/dm}
\ee

The hi-B regions could be primordial black holes, in particular quasars,
disperse clouds of antimatter, and unusual stars and anti-stars. All
may be not too far from us. These black holes may make all or a part
of the cosmological dark matter. Such dark matter is similar to normal
dark matter with one difference that the masses of DM ``particles'' are
not the same. Their dominant part may have masses close to the solar 
mass, but on the tail of the distribution very heavy black holes
of millions solar masses may exist. Since they are dispersed over mass,
their existence may be compatible
with the bounds on MACHOs in different mass intervals~\cite{macho}.

Primordial nucleosynthesis in hi-B domains proceeded with large
$ n_B/n_\gamma$ and the abundances of the early produced
elements would be quite different from the standard BBN~\cite{hi-b-bbn}.
In particular, primordial heavy nuclei and anti-nuclei, up to
iron or anti-iron could be formed. An early formation of heavy nuclei
may explain evolved chemistry around high redshift quasars.
If such hi-B regions were formed in our neighborhood, an observation
of heavy anti-nuclei might be plausible.

A different process of formation of high density baryonic bubbles was
proposed in ref.~\cite{zhit}. It was argued that during QCD phase transition
macroscopically heavy (with mass about $10^7$ g) objects might be formed 
which consisted either of baryons or antibaryons. Due to CP-violation there
could be a small excess of antibaryonic bubbles with respect to baryonic ones
and, vice versa, a small excess of free baryons over antibaryons. The latter
would be the observed cosmic baryons, while the former could make the cosmological
dark matter. Unfortunately no persuasive calculations have been presented 
(at least till now) to support this picture.

\section{Conclusion}

Universe existence clearly demonstrates that both CP and B are not 
conserved. Though models without CP and B breaking can be constructed,
they are much less natural.
Most probably CP-violation in cosmology is not directly related to CP-violation
observed in particle physics. Moreover, there exists a natural mechanism of
CP-breaking in the early universe which is untraceable today in direct
experiments. 

The standard scenarios of baryogenesis can explain
one number, the magnitude of the cosmological baryon asymmetry,
$\beta = const$. Due to this it is impossible to distinguish between them
and to understand what really happened in the early universe. 

There are two ways to resolve this uncertainty. First, to find a 
testable model of particle physics which allows to express the high energy 
parameters through the measurable low energy ones. A crucial demand
to such a model is the explanation of the close magnitudes of energy densities
of baryonic and dark matter. At the present stage it is hardly feasible.

Another possibility is to rely on cosmological good luck. If a non-standard
model of baryogenesis is realized, such that $\beta$ is not a constant
but a function of space points, $\beta = \beta (x)$, then measuring the
corresponding isocurvature perturbations we can hope to select the right
baryogenesis scenario. Especially interesting is the case if somewhere
$\beta <0$ i.e. cosmic antimatter is created (see more below).
In connection with ``good luck'' I would like to re-quote quotation from
Napoleon presented in one of the lectures by I. Bigi at this School: Napoleon
considered being lucky as a necessary condition for his generals. Maybe 
being lucky to exist, we also have some other portions of good luck, in
particular, to understand the universe.
Maybe anthropic principle includes good luck?

It is unknown if astronomically large domains of antimatter
exist not too far from us, though the universe is most probably
globally charge asymmetric at least inside the present day horizon volume.
Still some antimatter may be almost at hand with abundant heavy anti-nuclei.

Stochastic or dynamical CP-breaking leads to a very interesting pattern of
distribution of baryonic matter and antimatter. A search for isocurvature
fluctuations in the angular spectrum of CMBR and in large scale structure,
together with the search for cosmic antimatter and unusual sources of gamma
rays from matter-antimatter annihilation is of primary importance for
understanding the nature of the cosmological CP-violation.

As for more theoretical conclusions, we should mention that
despite breaking of T-invariance and related to it absence of detailed balance,
the canonical equilibrium distributions remain true.
Most probably they survive even if CPT is broken.

If unitarity and the canonical spin-statistics relation are broken then 
the usual equilibrium distributions may be broken too and the effects can be 
accumulated and large. Maybe stationary equilibrium distributions do not
exist. However,
one should be aware of Pandora's box of consequences if sacred principles 
are destroyed. I would like to finish with a quotation 
from ``Brothers Karamazov''
by Fedor Mihailovich Dostoevsky:
``If there is no God, everything is permitted.''


\end{document}